\documentclass[%
 reprint,
superscriptaddress,
 amsmath,amssymb,
 aps,
]{revtex4-1}
\newcommand{\vev}{\left<h\right>}
\newcommand\scalemath[2]{\scalebox{#1}{\mbox{\ensuremath{\displaystyle #2}}}}
\usepackage{enumitem}
\usepackage{amsmath}
\usepackage{slashed}
\usepackage{braket}
\usepackage{graphicx,subfigure}
\usepackage{dcolumn}
\usepackage{bm}
\usepackage{etoolbox}
\apptocmd{\thebibliography}{\raggedright}{}{}

\begin{document}

\preprint{APS/123-QED}

\title{Consequences of an Abelian $Z'$ for neutrino oscillations and dark matter}

\author{Ryan Plestid}
\email{plestird@mcmaster.ca}
%
%
 \affiliation{Physics Department, McMaster University, Hamilton, Ontario, Canada}
%
%
%

\date{\today}

\begin{abstract}
The Standard Model's accidental and anomaly-free currents, $B-L$, $L_e-L_\mu$, $L_e-L_\tau$, and $L_\mu-L_\tau$ could be indicative of a hidden gauge structure beyond the Standard Model. Additionally neutrino masses can be generated by a dimension-five operator that generically breaks all of these symmetries. It is therefore important to investigate the compatibility of a gauged $U'(1)$ and neutrino phenomenology. We consider gauging each of the symmetries above with a minimal extended matter content. This includes the $Z'$, an order parameter to break the $U'(1)$, and three right-handed neutrinos. We find all but $B-L$ require additional matter content to explain the measured neutrino oscillation parameters. We also discuss the compatibility of the measured neutrino textures with a non-thermal dark matter production mechanism involving the decay of the $Z'$. Finally we present a parametric relation that implies that any sterile neutrino dark matter candidate should not be expected to contribute to neutrino masses beyond ten parts per million.

\end{abstract}

\maketitle


\section{\label{sec:level1}Introduction}

In the past decade there has been considerable interest in the phenomenology of beyond the Standard Model, Abelian, gauge bosons that couple to leptons \cite{Baek:2008nz,Langacker:2008yv,Heeck:2011wj,Williams:2011qb,Harigaya:2013twa,Shuve:2014doa}. A new gauge boson would imply the existence of some new gauge symmetry structure. The Standard Model contains three global, independent, and accidental Abelian symmetries that are anomaly-free \cite{Foot:1990mn,He:1991qd}. From these one can form the combinations $B-L$,  $L_e-L_\mu$,  $L_\mu-L_\tau$, and $L_e-L_\tau$  \cite{WeinbergVol2}. Additionally, neutrino oscillations have been observed \cite{Ahmad:2002jz, Wendell:2010md, Abe:2010hy, Aharmim:2011yq, Zhan:2015aha} which are indicative of non-zero neutrino masses. This can be understood via the dimension-five Weinberg operator \cite{Weinberg:1979sa} that generically breaks all of the symmetries listed above. A theory which includes a $Z'$ coupled to one of these currents would constrain the form of the Weinberg operator (and by proxy the neutrino mass matrix). Thus it is worth considering the compatibility of these gauge symmetries with the measured oscillation data.  

We consider a model that extends the Standard Model gauge group by $G_{SM}\to G_{SM}\otimes U'(1)$ where this new $U'(1)$ will be associated with the aforementioned anomaly-free current that the $Z'$ is coupled to. Additional ingredients will also be included to induce neutrino masses, and to conform to bounds from $Z'$ phenomenology. 

Experimental and observational constraints for a new $Z'$  are dictated by the current to which it couples. Typically the strongest bounds come from  the  coupling to electrons. This is because the heavier flavour counterparts are unstable, and therefore beam dump and solar neutrino absorption experiments involve electrons, and electron neutrinos respectively \cite{D'Angelo:2014vgk, Bjorken:2009mm, Essig:2009nc}. In the case that there is no such coupling at tree-level (i.e. $L_\mu-L_\tau$), the processes probed by these experiments are mediated by loops and can be sub-dominant to other constraints arising from other sources, such as neutrino-trident production \cite{Altmannshofer:2014pba}.  
%

A $Z'$ that couples to any of the lepton generations must satisfy bounds from from measurements of the cosmic microwave background that constrain the number of effective relativistic degrees of freedom $N_{eff}$ \cite{Lesgourgues:2006nd}; this is related to the decay $Z'\to\nu+\nu$. If the $Z'$ decays when the temperature of the universe is approximately $1~\text{MeV}$, then the resultant spike in the neutrino population will not have sufficient time to return to its equilibrium distribution before freeze-out. This sets a bound on the mass of the $Z'$ of roughly $M_{Z'}>4~\text{MeV}$ \cite{Ade:2015xua}. A massive $Z'$ implies that the gauge symmetry must be spontaneously broken. 
%
%
%
%

As previously mentioned, an interesting place to search for the imprints of these possible gauge symmetries is in the neutrino oscillation data collected over the past fifteen years \cite{Ahmad:2002jz,Wendell:2010md,Abe:2010hy,Zhan:2015aha}. This can give us information about the PMNS matrix, which dictates lepton-flavour-violating processes \cite{Pontecorvo:1967fh}. These oscillations are the result of neutrinos having a non-zero mass. In the context of the Standard Model these can be induced by adding additional fermions that do not couple to any of the known forces \cite{Kusenko:2009up}; these are often termed right-handed, or sterile, neutrinos.

Beyond explaining neutrino oscillations, if there were three of these right-handed neutrinos the Standard Model would be significantly more symmetric in the sense that each left-handed fermion would have a corresponding right-handed fermion. This final statement assumes that the three right-handed neutrinos would be labelled by  the conventional generation indices $\{e,\mu,\tau\}$. We would like to investigate this minimal, and aesthetically attractive, extension of the Standard Model in conjunction with a new $U'(1)$ gauge symmetry and its associated $Z'$. 

To ensure that bounds on the mass of a lepton-coupled $Z'$ are satisfied ($M_{Z'}>4~\text{MeV}$ \cite{Ade:2015xua}), a scalar field, $S$, charged under the $U'(1)$ is included to spontaneously break the gauge symmetry. Additionally we extended the Standard Model by including three standard model singlets: $N_e$, $N_\mu$, and $N_\tau$. These fields carry non-zero charge under $L_e$, $L_\mu$, and $L_\tau$ respectively and are totally decoupled from the Standard Model gauge bosons. We will refer to this  model as the $3N$-extension when only renormalizable operators are included (this is similar to the $\nu$MSM \cite{Asaka:2005pn,Asaka:2005an} but with an extended gauge group). 

In the case of $B-L$ the three right-handed neutrinos are required by anomaly cancellation \cite{Foot:1990mn}. For the lepton flavour symmetries ($L_e-L_\mu$, $L_\mu-L_\tau$, and $L_e-L_\tau$) these additional fields are not required if one gauges only one of currents, but are motivated by observed neutrino oscillations and the previously mentioned aesthetics. If one wished to gauge two of these symmetries simultaneously, one would be forced to include these additional right-handed states \cite{Foot:1990mn, He:1991qd} to remove anomalies. 
%


Right-handed neutrinos also lend themselves to being a natural dark matter candidate for masses in the range of $1-100~\text{keV}$ \cite{Dodelson:1993je, Dolgov:2000ew, Wu:2009yr}. The original proposal for right-handed neutrinos as a dark matter candidate was made by Dodelson and Widrow \cite{Dodelson:1993je}. It relied on a non-thermal production mechanism mediated by Standard Model physics, in which sterile mass eigenstates were produced via the weak mass-mixing of right-handed and left-handed neutrinos. Galactic x-ray searches \cite{Boyarsky:2006ag, Watson:2012yr} and small scale structure formation \cite{Shuve:2014doa} have since excluded the viable parameter space for the Dodelson-Widrow proposal. 

Although the Dodelson-Widrow scenario has been excluded, there exist ``Dodelson-Widrow-esque'' proposals that are still viable \cite{Kusenko:2009up}. One possible way to satisfy the bounds mentioned above is to have the dark matter be generated by some beyond the Standard Model process, such as the $Z'$  progenitor scenario proposed by Shuve and Yavin \cite{Shuve:2014doa}. Here the sterile neutrino dark matter abundance is generated via the decay of a massive $Z'$, and never comes into thermal equilibrium with the photon bath. The $Z'$ only couples to the dark matter via mass-mixing of the dark matter and some set of left-handed Standard Model neutrinos. This scenario was found to be viable for a very weakly coupled ($g'\thicksim 10^{-3}-10^{-6}$), and massive ($M_{Z'}\thicksim \text{MeV}-\text{GeV}$) $Z'$ with mass-mixing characterized by a mixing angle $\theta$ defined, at zero temperature, by
%
%
\begin{equation}
\left[\begin{array}{c}
\nu_1\\
\nu_2 
\end{array}\right]=
\left[\begin{array}{cc}
\cos{\theta} & -\sin{\theta}\\
\sin{\theta} & \cos{\theta}
\end{array}\right]
\left[\begin{array}{c}
\nu_a\\
N_s
\end{array}\right]\label{mixingAngle}
\end{equation}
where $\nu_a$ and $N_s$ are the active and sterile states in the flavour basis with respect to $Z'$. The fields  $\nu_1$ and $\nu_2$ are the ``mostly active'' and ``mostly sterile'' mass eigenstates respectively, with the understanding that $\theta\leq \pi/4$.

The $Z'$ must couple to some anomaly-free current involving lepton number for this scenario to be viable. Shuve and Yavin considered the case of $L_\mu-L_\tau$, however they mention the dark matter production mechanism is still viable with other currents, such as $B-L$ or $L_e-L_\mu$. The Shuve-Yavin progenitor scenario relies on indirect coupling to the dark matter via a mixing angle. This means that, of the added right-handed neutrinos, at least one must be decoupled from the $Z'$.

The $3N$-extension discussed earlier has one right-handed neutrino for each lepton generation in the Standard Model. In the case of a gauged $B-L$ symmetry, this provides no ``sterile state'' since all of the the right-handed neutrinos would couple to the $Z'$ because they all carry lepton number. Anomaly-free lepton flavour symmetries that only couple to two lepton generations, such as $L_\mu-L_\tau$, naturally yield a single ``sterile state''; for $L_\mu-L_\tau$ it is $N_e$. 

Some natural questions to ask are: can our $3N$-extension account for both the observed dark matter abundance, and the neutrino oscillation data? Can oscillation data shed any light on our picture of sterile neutrino dark matter?

In Section \ref{sec:2} we discuss the effects of the choice of gauge symmetry for both neutrino masses and dark matter. In Section \ref{sec:3} we attempt to produce the neutrino textures in a model with a gauged $L_\mu-L_\tau$. We first attempt to use just the $3N$-extension, and then subsequently investigate the effects of including higher-dimensional operators.  In Section \ref{sec:ImplicNeutrino} we discuss the implications of neutrino oscillation data for the $Z'$ progenitor scenario. Section \ref{sec:conc} contains a summary of our findings. 

Those who are primarily interested in neutrino textures should consult Sections \ref{naivety} and \ref{sec:3}. Those primarily interested in sterile neutrino dark matter should focus on Sections \ref{progenitorOver}, and \ref{sec:ImplicNeutrino}, but it should be noted these sections use results from the neutrino texture discussion. 

\section{The Naive Consequences of One's Choice of Current}\label{sec:2}
\subsection{Neutrino phenomenology}\label{naivety}
\subsubsection{$B-L$ as the gauge symmetry}
Let us consider $B-L$ first for concreteness. In this case all of the right-handed fields carry the same charge. In the absence of a coupling to an order parameter, this implies that all Majorana masses must vanish so that neutrinos would be Dirac particles. The Yukawa coupling matrix is defined by the contribution to the Lagrangian $Y_{ij}L_i\tilde{H}N_j$ where $\tilde{H}\equiv \epsilon H^{\dagger}$, $H$ is the Higgs doublet, and $L$ is the Lepton doublet. For an unbroken $B-L$ symmetry this matrix is unconstrained and dictates the mixing exclusively. In this case, neutrino phenomenology can be obtained trivially due to the number of degrees of freedom present in the Yukawa matrix. 

Due to the fact that all of the right-handed neutrinos carry charge under $B-L$ any Majorana mass term is forbidden since the combination $N_i N_j$ is not invariant under the gauge group. Yukawa couplings are allowed since $N_i$ carries charge opposite to $L_j$ under $U_{B-L}(1)$ and so $Y_{ij}L_i\tilde{H}N_j$ is a gauge singlet. Since the gauge symmetry is flavour blind all possible entries in the Yukawa matrix are allowed and all nine of its independent entries are populated.  
\begin{equation}
Y=\left[\begin{array}{ccc}
\times & \times & \times\\
\times & \times & \times \\
\times & \times & \times \end{array}\right]~~~
M_{R}=\left[\begin{array}{ccc}
0 & 0 & 0 \\
0 & 0  & 0 \\
0 & 0 & 0 \end{array}\right]
\end{equation}

Next we consider couplings involving the complex order paramter $S$. With appropriate charge assignments (allowing terms like $\Gamma_{ij}SN_iN_j\to M_{ij}N_i N_j$) the spontaneous breaking of the symmetry by the scalar field $S$ can result in a fully populated mass matrix. 
\begin{equation}
Y=\left[\begin{array}{ccc}
\times & \times & \times\\
\times & \times & \times \\
\times & \times & \times \end{array}\right]~~~
M_{R}=\left[\begin{array}{ccc}
\times & \times & \times\\
\cdot & \times & \times \\
\cdot & \cdot & \times \end{array}\right]
\end{equation}
Here $Y$ is the Yukawa matrix, $M_R$ is the right handed mass matrix, $\times$ denotes an independent entry, and $\cdot$ denotes an entry determined by the required symmetry of $M_R$. There are clearly enough degrees of freedom in $B-L$ to achieve the correct neutrino phenomenology. Even for Dirac neutrinos (i.e. $M_R=0$), in the $CP$ conserving limit the Yukawa matrix has nine free parameters.

\subsubsection{$L_\mu-L_\tau$ as the gauge symmetry}
In contrast to $B-L$ it is not obvious that the correct neutrino phenomenology can be recovered from the model in the case of $L_\mu-L_\tau$. This is because the generational dependence of the symmetry restricts the form of the Yukawa matrix. There are additional degrees of freedom in the context of neutrino mixing due to the presence of mass terms that allow for the coupling of right handed fields to one another. Specifically the Lagrangian may contain mass terms coupling $N_\mu$ and $N_\tau$ as a Dirac pair while $N_e$ may have a Majorana mass. In this case we get
\begin{equation}
Y=\left[\begin{array}{ccc}
\times & 0 & 0\\
0 & \times & 0\\
0 & 0 & \times \end{array}\right]~~~
M_{R}=\left[\begin{array}{ccc}
\times & 0 & 0 \\
0 & 0  & \times \\
0 & \cdot & 0 \end{array}\right]
\end{equation}

In contrast to $B-L$, $L_\mu-L_\tau$ constrains the Yukawa matrix so severely that the mixing is controlled primarily by the right-handed mass matrix. In the case of an unbroken $L_\mu-L_\tau$ this structure gives a maximal $\theta_{23}$ and vanishing mixing with the first generation ($\theta_{12}=\theta_{13}=0$). 

With the inclusion of an order parameter that breaks the gauge symmetry additional terms in the mass matrix are made possible. $B-L$ is generation blind and so, as we saw in the previous section, if the order parameter generates any mass terms in the broken phase it will generically populate the entire mass matrix. By contrast, due to the structure of $L_\mu-L_\tau$ the charge of the order parameter dictates which entries in the right-handed mass matrix will be non-zero. Let us set the relative charges of the right-handed fields, and the order parameter by $Q(S)=Q(N_\tau)=-Q(N_\mu)$. Then we get 
\begin{equation}
Y=\left[\begin{array}{ccc}
\times & 0 & 0\\
0 & \times & 0\\
0 & 0 & \times \end{array}\right]~~~
M_{R}=\left[\begin{array}{ccc}
\times & \times & \times\\
\cdot & 0 & \times \\
\cdot & \cdot & 0 \end{array}\right]\label{B-Lmass}
\end{equation}
up to renormalizable operators. We see that there are seven independent parameters, as well as mixing between the first, second, and third generations. Neutrino oscillations are controlled by the active-neutrino mass matrix, $M_L$, which in the see saw limit is given by $Y^TM_R^{-1}Y$. There are only five oscillation parameters that have been measured precisely and so we naively expect that this set up has enough degrees of freedom (7) to fit to the neutrino oscillations parameters (5). This will be discussed in greater detail in Section \ref{sec:3}.

The most important distinction between $B-L$, and $L_\mu-L_\tau$ is that the former allows for a totally unconstrained set of right-handed mixing parameters while the latter reduces the number of free parameters considerably. This means that while a theory of gauged $B-L$ will certainly be able to fit the neutrino oscillation data, it will not be able to give an explanation for its structure beyond arbitrary choices of parameters.

\subsection{Shuve-Yavin progenitor scenario}\label{progenitorOver}
We will begin by reviewing the necessary ingredients for the Shuve-Yavin progenitor scenario to represent a viable dark matter production mechanism. The scenario relies on mass-mixing between a totally sterile dark matter candidate and an active state \cite{Shuve:2014doa}. This is necessary to allow for a non-thermal freeze-in scenario where high temperature effects strongly suppress the mixing in the early universe \cite{Notzold:1987ik}. For this to occur we need to include a state which is totally sterile with respect to the Standard Model and the $Z'$. This totally sterile state must have some mass mixing, with some set of states, that couple to the $Z'$; these states can be both left- and/or right-handed. This mixing must be sufficiently small so as to avoid bounds from galactic x-ray searches \cite{Watson:2006qb, Boyarsky:2006ag, Watson:2012yr, Horiuchi:2013noa}. Not only does this state need to satisfy astrophysical bounds it must be stable to ensure that the dark matter survives from early epochs to today. For this to occur we need the sterile state to be the lightest right-handed mass eigenstate. Otherwise the decay $N_s\to N_a+\nu+\nu$ is viable, where $N_a$ is a lighter mass eigenstate that couples to the $Z'$. 

So to employ the production mechanism envisioned by Shuve and Yavin we must:
\begin{enumerate}[label=(\alph*)]
\item Include at least one sterile neutrino that is totally uncharged under the gauge symmetry.

\item Ensure the smallest eigenvalue of the right-handed mass matrix corresponds to the mostly weakly coupled eigenstate mentioned above.
\end{enumerate}
$L_\mu-L_\tau$ gives us a sterile state for free in the $3N$-extension. $N_e$ has vanishing charge and so we would naturally identify some mass eigenstate that is ``mostly-$N_e$'' as the ``mostly-sterile'' eigenstate; this is the dark matter candidate. Since we expect this state to be involved in the production of neutrino masses the necessary parameter choices for viable  neutrino phenomenology would have to be consistent with those of the Shuve-Yavin progenitor scenario. 

$B-L$ does not yield a totally sterile state in the $3N$-extension, because three of the right-handed neutrinos must be charged under $B-L$ (due to anomaly cancellation requirements). Thus to employ the Shuve-Yavin mechanism for a $Z'$ coupled to the current $J^\mu_{B-L}$ we would need to introduce an additional Standard Model singlet, $N'$. The mixing of this state with any active states is forbidden in the absence of an order parameter. With an order parameter included mass-mixing is dictated by our charge assignments. We must enforce $Q(S)=-2Q(N_i)$ to generate a non-vanishing right handed mass matrix  like the one shown in  Equation \ref{B-Lmass}. This assignment forbids the existence of mass-mixing between the $N'$ and the three right-handed neutrinos. We can only write down the Majorana mass term $\frac{1}{2}M'N'N'$. If we instead choose $Q(S)=-Q(N_i)$ then we may write $y_iSN_i N'$. After the order parameter breaks the symmetry this will result in a Dirac mass that mixes $N'$ with the active right-handed states $N_i$, however the right-handed mass matrix will not be populated. Thus a $B-L$ Shuve-Yavin progenitor scenario would necessarily involve an additional sterile-state $N'$ and Dirac neutrinos. This seems less attractive then the case of $L_\mu-L_\tau$ which, at the level of naive qualitative analysis, seems to work for a $3N$-extension.  
%
%

So we see that if our $Z'$ couples to $B-L$ we must introduce a right-handed state that can not be classified as belonging to one of the three existing lepton families. By contrast, at the level of qualitative analysis a gauged $L_\mu-L_\tau$ seems like it may be capable of yielding a dark matter candidate, and appropriate neutrino phenomenology, in the 3N-extension.

\section{Neutrino Phenomenology for a Gauged $L_\mu-L_\tau$}\label{sec:3}
\subsection{A renormalizable theory}
Having examined the qualitative features of mass-mixing in our model in Section \ref{sec:2} we now wish to investigate the model's neutrino textures quantitatively. As was discussed in the introduction our model extends the Standard Model using three right handed neutrinos labelled by the standard lepton generational indices and a $Z'$ coupled to the charge $L_\mu-L_\tau$. This symmetry is spontaneously broken by an order parameter, $S$, that transforms as a singlet under the Standard Model gauge group but is charged under the newly gauged, accidental $U'(1)$ symmetry. This model will be defined by Lagrangian
%
%
\begin{equation}
\mathcal{L}=\mathcal{L}_{SM}+\mathcal{L}_{Z'}
\end{equation}
where $\mathcal{L}_{SM}$ is the Standard Model Lagrangian and $\mathcal{L}_{Z'}$ contains that beyond Standard Model physics. This new physics dictates the interactions of our three new right-handed fields, $N_e$, $N_\mu$, and $N_\tau$, with the Standard Model, the new $Z'$, and the symmetry breaking order parameter $S$. The charge assignments will be such that $Q(L_\mu)=-Q(N_\mu)=Q(S)=Q(N_\tau)=-Q(L_\tau)$ where $L_i$ are the familiar Standard Model doublets. The Standard Model singlets $N_\mu$ and $N_\tau$ have the same charges as $N_\mu$ and $N_\tau$ respectively. This leads to the following Yukawa and right-handed mass matrices \cite{Choubey:2004hn,Araki:2012ip}
\begin{equation}
Y=\left[\begin{array}{ccc}
y_e & 0 & 0\\
0 & y_\mu & 0\\
0 & 0 & y_\tau \end{array}\right]~~~
M_{R}=\left[\begin{array}{ccc}
M_e & m_\mu & m_\tau\\
\cdot & 0 & M \\
\cdot & \cdot & 0 \end{array}\right]\label{rightHandedMassMatrix1}
\end{equation}
where $m_\mu$ and $m_\tau$ are generated by the spontaneous breaking of the $U'(1)$ by the field $S$. As was previously discussed this seems to suggest that neutrino mixing will be controlled by seven parameters. However, assuming that we may work in the see-saw limit and applying the see-saw relation $M_L=m_D M_R^{-1} m_D^T$ we obtain
\begin{equation}\scalemath{1.2}{
M^{(L)}=\overline{\mathcal{M}}\left[
\begin{array}{ccc}
1 &-\frac{r}{\mu} & -\frac{r^{-1}}{\mu} \\
\cdot  & \frac{r^2}{\mu^2}& \frac{\mu\cdot \mu_e-1}{\mu^2}\\
\cdot & \cdot &  \frac{r^{-2}}{\mu^2}
\end{array}\right]}
\end{equation} \label{paramOfMassMatrix}
These variables are related to the original set of parameters by 
\begin{subequations}\label{Vanilladefinitions}
\begin{equation}
r\equiv\sqrt{\frac{m_\tau y_\mu}{y_\tau m_\mu}}
\end{equation}
\begin{equation}
\mu\equiv\frac{y_e M}{\sqrt{y_\mu y_\tau}\sqrt{m_\mu m_\tau}}
\end{equation}
\begin{equation}
\mu_e\equiv\frac{\sqrt{y_\mu y_\tau} M_e}{y_e\sqrt{m_\mu m_\tau}}
\end{equation}
\begin{equation}
\overline{\mathcal{M}}\equiv\frac{y_e\sqrt{ y_\mu y_\tau}\vev^2}{\sqrt{m_\mu m_\tau }}\frac{\mu }{\mu\mu_e-2}
\end{equation}
\end{subequations}
We can immediately see that in fact the left-handed mass matrix, and consequently neutrino oscillation data, is controlled by only four parameters. This is a consequence of the diagonal Yukawa matrix and the fact that our left handed mass matrix contained only four entries. If we define $\Omega_{ij}\equiv \frac{1}{M_{11}}M_{ij}$ then we can see that parametrization of the symmetric $3\times3$ matrix by only four parameters manifests itself in the relations $\Omega_{12}^2=\Omega_{22}$ and $\Omega_{13}^2=\Omega_{33}$. If some choice of these four parameters were to be able to produce a neutrino texture that agreed with observations this would be suggestive evidence that neutrino textures depend on some $\mu,\tau$ flavour symmetries. 

This is because, if we consider the $CP$ conserving limit, neutrino textures are controlled by the set of six parameters $\{m_1,m_2,m_3,\theta_{12},\theta_{13},\theta_{23}\}$. The parameters that have been measured are $\{\Delta m^2_{12},\Delta m_{h\ell}^2,\theta_{12},\theta_{13},\theta_{23}\}$ where $\Delta m^2_{12}\equiv m^2_2-m^2_1$ and $\Delta m_{h\ell}^2$ is defined similarly but for the heaviest and lightest eigenstates. This is equivalent to five of six possible pieces of information. Our matrix contains only four free parameters including an overall mass scale. If we could fit to the neutrino textures our model would actually be predictive. We would only need to use four of the experimental quantities to predict the entire texture.

It can be shown \cite{plestid:2015thesis} that this minimal model is incapable of reproducing the neutrino data to within $3\sigma$ for all of the central fit values from nu-Fit \cite{Gonzalez-Garcia:2014bfa}. However good qualitative agreement is found with the model capable of fitting to four of the five measured values within $2\sigma$ . The solution that fits all of the parameters except $\theta_{12}$ within $2\sigma$ is about $25\%$ off the central fit value for $\sin^2{\theta_{12}}$. Due to the increased precision in neutrino experiments this represents a $4.4\sigma$ deviation from the best-fit values. The solutions that fit all parameters except $\theta_{13}$ within $2\sigma$ predict a near vanishing $\sin^2{\theta_{13}}$. 

The ability of our $3N$-extension to produce the correct qualitative features of the neutrino textures is interesting and suggests that perhaps the correct neutrino phenomenology can be obtained via some small perturbation on this minimal model. In the following section we consider the effects of higher dimensional operators on the model's ability to fit to the observed neutrino oscillation data, and see if these introduce enough degrees of freedom to fit to the observed textures. 

\subsection{Including higher-dimensional operators}\label{sec:HighDim}
The simplest solution that allows this model to fit to the measured neutrino parameters is to add additional degrees of freedom in the form of higher-dimensional operators. These operators will be most easily realized in a UV completion by adding additional heavy sterile states to the theory which can be integrated out. By determining what dimension-five operators we need to add to the model we can gain some insight about possible UV completions.

The types of operators we will consider will broadly fall into two categories: those that affect the Yukawa matrix entries (Yukawa-inducing operators) and those that affect the mass matrix (mass-inducing operators). 

\subsubsection{Mass-Inducing Operators}
The three, dimension-five, mass-inducing operators we can write down are
\begin{subequations}
\begin{equation}
\mathcal{L}_{Z'}\supset \frac{\delta M_\mu}{\left<\sigma\right>^2}S^2N_\mu N_\mu
\end{equation}
\begin{equation}
\mathcal{L}_{Z'}\supset \frac{\delta M_\tau}{\left<\sigma\right>^2}\left(S^\dagger\right)^2N_\tau N_\tau
\end{equation}
\begin{equation}
\mathcal{L}_{Z'}\supset \frac{\delta M_e}{\left<\sigma\right>^2}S^\dagger S N_e N_e
\end{equation}
\end{subequations}
where the $\delta M_i$ are the resulting contributions to the right-handed mass matrix after $U'(1)$ symmetry breaking. The field $\sigma$ is the radial component of the order parameter defined by $S\equiv\frac{1}{\sqrt{2}}\sigma \exp{\left[i \pi/\left<\sigma\right>\right]}$ and $\left<\sigma\right>$ is this fields vacuum expectation value.  After symmetry breaking we may make the replacement $S\to \frac{1}{\sqrt{2}}\left<\sigma\right>$which induces mass terms in the diagonal entries of $M_R$. The third of these operators is not of interest to us because its effects are equivalent to a redefinition of $M_e$. 

This yields a right handed mass matrix given by
\begin{equation}
M_R=\left[\begin{array}{ccc}
M_e & m_\mu & m_\tau\\
\cdot & \delta M_\mu & M \\
\cdot & \cdot & \delta M_\tau 
\end{array}\right]
\end{equation}
with the same Yukawa matrix as before. 

There is of course no reason to expect \emph{a priori} that $\delta M_\tau$ is significantly smaller than $\delta M_\mu$, and there is no issue if they are comparable in size. This just gives six degrees of freedom in the $CP$ conserving limit. This is enough to parametrize any general $3\times3$ symmetric matrix, and so with this combination any neutrino texture could be generated. This is contrary to the aesthetics that initially directed us towards studying flavour dependent currents over the flavour blind $B-L$. It turns out that there exist solutions with $\delta M_\tau=0$ and so for the sake of simplicity we will consider this limit.
\begin{equation}
M_L=\overline{\mathcal{M}}\left[
\begin{array}{ccc}
1 &-\frac{r}{\mu} & -\frac{r^{-1}}{\mu}+\delta\frac{r}{\mu} \\
\cdot & \frac{r^2}{\mu^2}& \frac{\mu\cdot \mu_e-1}{\mu^2}\\
\cdot & \cdot &  \frac{r^{-2}}{\mu^2}-\delta \frac{\mu_e}{\mu}
\end{array}\right]\label{MASSMATRIXCHAP4}
\end{equation}
These parameters are the same as those introduced in Equation \ref{Vanilladefinitions} with the two new definitions,
\begin{subequations}
\begin{equation}
\overline{\mathcal{M}}\equiv\frac{y_e\sqrt{ y_\mu y_\tau}\vev^2}{\sqrt{m_\mu m_\tau }}\frac{\mu }{\mu\mu_e-2+\delta r^2}
\end{equation}
\begin{equation}
\delta\equiv\frac{\delta M_\mu y_\tau}{M y_\mu}
\end{equation}
\end{subequations}  
In the limit that $\delta\to0$ we recover the results for the renormalizable case. The key here is that if we define $\Omega$ as before we still have that $\Omega_{12}^2=\Omega_{22}$ but we no longer have the condition $\Omega_{13}^2=\Omega_{33}$.  

With this extra degree of freedom a finite number of solutions can be found.  For example $\{\mu,\mu_e,\delta,r \}=\{ -0.741,0.641, -0.380, 0.976\}$ yields a neutrino texture that agrees with the central values from nu-Fit \cite{Gonzalez-Garcia:2014bfa}. This solution exists very close to the  $\mu\leftrightarrow\tau$ exchange symmetry in the $3N$-model ($r=1$) with the $\delta$ term allowing for the accommodation of a non-zero $\theta_{13}$ by explicitly breaking this symmetry. 

All of the solutions that we found generically predict $\delta$ to be $\mathcal{O}(1)$. This implies for $y_\mu\thicksim y_\tau$ that $\delta M_\mu \thicksim M$. This does not contradict our expectation that this operator is generated at tree level by a heavy right-handed state. Suppose we added a state $N'$ with Majorana mass $M'$ that coupled to $S$ via a Yukawa interaction like $gSN'N_\mu$. Then we would expect $\delta M_\mu \thicksim g \left<\sigma\right> \frac{g\left<\sigma\right>}{M'}$. Provided $M'\gg g\left<\sigma\right> \gg M$ (i.e. a see-saw hierarchy) and that we have the coincidence of scales that
\begin{equation}
\frac{g\left<\sigma\right>}{M'}\thicksim \frac{M}{g\left<\sigma\right>}\label{coincidence}
\end{equation}
then $\delta M_\mu \thicksim M$ is perfectly consistent with an effective field theory picture.

\subsubsection{Yukawa-Inducing Operators}
A second class of operators we could include are those which would affect the Yukawa matrix. These operators will be of the schematic form $L_i\tilde{H} N_j S $. More specifically we can write down four possible operators
\begin{subequations}
\begin{equation}
\mathcal{L}_{Z'}\supset \frac{Z_\mu}{\tfrac{1}{\sqrt{2}}\left<\sigma\right>}S L_e \tilde{H}N_\mu
\end{equation}
\begin{equation}
\mathcal{L}_{Z'}\supset \frac{Z_\tau}{\tfrac{1}{\sqrt{2}}\left<\sigma\right>}S^\dagger L_e\tilde{H} N_\tau 
\end{equation}
\begin{equation}
\mathcal{L}_{Z'}\supset \frac{\chi_\mu}{\tfrac{1}{\sqrt{2}}\left<\sigma\right>}S^\dagger L_\mu \tilde{H} N_e 
\end{equation}
\begin{equation}
\mathcal{L}_{Z'}\supset \frac{\chi_\tau}{\tfrac{1}{\sqrt{2}}\left<\sigma\right>}S L_\tau \tilde{H} N_e 
\end{equation}
\end{subequations}

We were originally motivated to study higher-dimension operators to gain intuition about possible UV completions. In the case of the mass-inducing operators we found a UV completion with only one additional sterile state was sufficient to generate the necessary dimension five operators. This makes the investigation of Yukawa-inducing operators seem unnecessary, however these operators are of interest if one wishes to take advantage of the production mechanism proposed by Shuve and Yavin.  One may expect that x-ray constraints would require $N_e$ to decouple (i.e. $y_e\thicksim 0$). In this limit the additional degrees of freedom afforded by the operators $Z_\mu$ and $Z_\tau$  still allow for the correct neutrino phenomenology to be generated.  This will be discussed further in Section \ref{LmuDiscussion}.

Since we wish to study a decoupled $N_e$, we only consider the $Z_{\mu,\tau}$ operators (i.e. we assumed $\chi_\mu=\chi_\tau=0$). We were able to show that with vanishing mass-inducing operators we were still able to obtain the correct neutrino phenomenology. For this to work we found that $Z_{\mu,\tau}\thicksim y_{\mu,\tau}$. This would require a coincidence of scales and parametrics similar to that found in Equation \ref{coincidence}.

\section{The Implications of Neutrino Physics for the Progenitor Scenario}\label{sec:ImplicNeutrino}
\subsection{General considerations for sterile neutrino dark matter}\label{DMandneutrinos}
\begin{figure}[t]
\centering     
\subfigure
{\label{fig:a}\includegraphics[width=40mm]{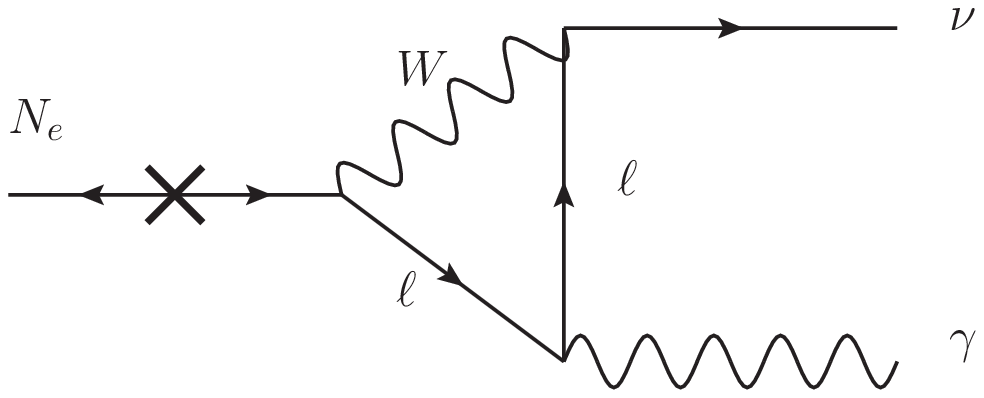}}
\subfigure{\label{fig:b}\includegraphics[width=40mm]{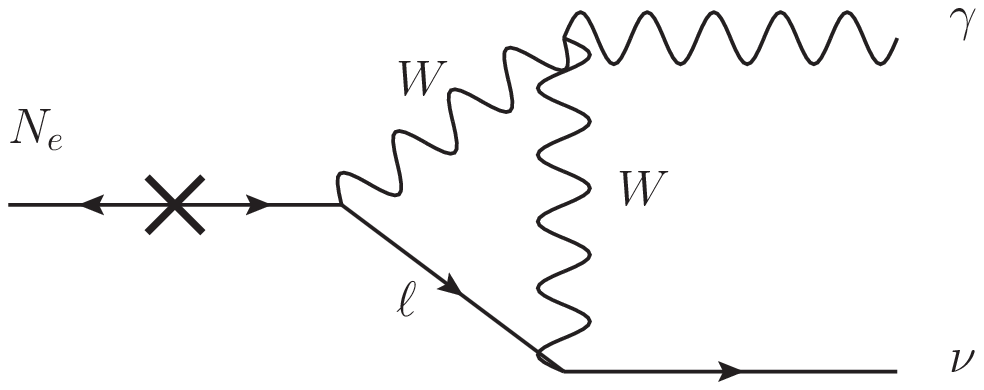}}
\caption{The process $N_s\to \nu + \gamma$ proceeding via a loop of W-bosons and leptons. The decay rate is proportional to $\sin^2{\theta}$.}\label{Steriledecay}
\end{figure}

If one assumes that some $\mathcal{O}(1)$ fraction of the dark matter is composed of sterile neutrinos, then one generically predicts the existence of galactic x-rays \cite{Kusenko:2009up}. This is because a sterile neutrino should have some coupling to active neutrinos parameterized by the mixing angle in Equation \ref{mixingAngle}, which in turn allows for the decay shown in Figure \ref{Steriledecay}. 

Galactic x-ray searches have constrained the mixing angle for various dark matter masses. In the case of a $10~\text{keV}$ dark matter candidate the bound can be conservatively stated as $\sin^2{\theta}<10^{-11}$ \cite{Horiuchi:2013noa}. It is worth discussing whether or not our parameter-fit to neutrino data remain viable after including this constraint.

We first consider the requirements of the $Z'$ scenario. An important feature is that the lightest mass eigenstate must be composed primarily of a totally sterile state with some very small fraction of active states (see Section \ref{progenitorOver}).

In the case of $L_\mu-L_\tau$ this eigenstate must be primarily $N_e$. Any mixing between $N_e$ and either $N_\mu$ or $N_\tau$ will result in a coupling between the lightest eigenstate $N_s$ and the active leptons. Additionally a non-zero $y_e$ will lead to mixing with first generation Standard Model leptons. 

In principle there is more than just one mixing angle to keep track of because we are discussing the mass eigenstates formed by various combinations of six fields. To a very good approximation we can reduce this to a two state problem by the following procedure.
\begin{enumerate}[label=(\roman*)]
\item Diagonalize the right-handed mass matrix. Identify $N_s$.
\item Find the linear combination of left-handed neutrinos that couple to $N_s$. Call this state $\nu_{\text{\tiny{$\Sigma$}}}$.
\item Integrate out the two heavier eigenstates $N_1$ and $N_2$. This will imbue $\nu_{\text{\tiny{$\Sigma$}}}$ with a mass term via the see-saw mechanism. 
\end{enumerate}
This will leave us with the following mass matrix to diagonalize 
\begin{equation}
\left[\begin{array}{cc}
m_{\text{\tiny{$\Sigma$}}} & y_{\text{\tiny{$\Sigma$}}}v \\
y_{\text{\tiny{$\Sigma$}}}v & M_s
\end{array}\right]
\end{equation}
where $M_s$ is the mass of the dark matter candidate, $y_{\text{\tiny{$\Sigma$}}}$ is the effective Yukawa coupling of the combination $N_s$ and $\nu_{\text{\tiny{$\Sigma$}}}$ to the Higgs field, and $v$ is the Higgs' vacuum expectation value. 

We know that $m_\nu\thicksim 0.1~\text{eV}$ and that right-handed neutrinos are good dark matter candidates for $M_s\thicksim 1~\text{keV}- 100~\text{keV}$. Therefore $m_{\text{\tiny{$\Sigma$}}}$ is negligible if we wish to consider a scenario involving sterile neutrino dark matter. This leads to the familiar see-saw relationships
\begin{subequations}
\begin{equation}
\theta= \frac{y_{\text{\tiny{$\Sigma$}}}v}{M_s}
\end{equation}
\begin{equation}
m_\nu^{(s)}=\frac{\left(y_{\text{\tiny{$\Sigma$}}}v\right)^2}{M_s}
\end{equation}
\end{subequations}
where $m_\nu^{(s)}$ is $N_s$'s contribution to the neutrino masses. This implies that 
\begin{equation}
m_\nu^{(s)}=M_s \theta^2_0 =M_s\sin^2{\theta}.
\end{equation}
So using the data from Figure 4 of ref. \cite{Horiuchi:2013noa} we see that  $\sin^2{\theta}<10^{-11}$ for $10~\text{keV}$ dark matter. This implies that
\begin{equation}
m_\nu^{(s)}< 10^{-9}~\text{keV}\thicksim 10^{-5} m_\nu
\end{equation}
where $m_\nu$ is the neutrino mass scale. This suggests that a sterile neutrino dark matter candidate will not contribute to neutrino masses beyond the level of about ten parts per million. Thus, sterile neutrinos that are eligible dark matter candidates should be thought of as entirely distinct from sterile neutrinos that are to explain neutrino oscillations. 

\subsection{The case of a gauged $L_\mu-L_\tau$\label{LmuDiscussion}}
In Section \ref{sec:HighDim} found that with the addition of dimension-five operators (only one was required for the mass-inducing operators) neutrino phenomenology can be accurately produced. The mass-inducing operator solution fixes the value of five parameters. Many of these turn out to be proportional to $y_e$, $m_\mu$, and $m_\tau$; parameters we know will lead to coupling with active states. The argument given above suggests that if $N_e$ is to be considered the dark matter candidate we must be able to obtain the correct neutrino phenomenology in the limit of a totally decoupled $N_e$. So we should consider how to obtain the correct phenomenology with only two right-handed neutrino species coupled to the Standard Model $N_\mu$ and $N_\tau$. 

Dimension-five operators were investigated and for non-vanishing $Z_\mu$, $Z_\tau$, $\delta M_\mu$ and $\delta M_\tau$ we could accommodate the appropriate neutrino textures. Including all four of these operators is equivalent to adding two additional heavy sterile states which are subsequently integrated out; this can be understood easily. 

Let us add only one additional field $N'$ which we would like to serve as the dark matter candidate. We are forced to totally decouple $N'$ to a first approximation so that it does not contribute to neutrino masses as argued in Section \ref{DMandneutrinos}. This reduces to our theory with $N_e$, $N_\mu$, and $N_\tau$ generating all of the neutrino phenomenology which was found to be incompatible with the observed neutrino textures. Thus we must add a second sterile state $N'_e$. This state, along with a heavy $N_e$ would be responsible for generating the appropriate Yukawa-inducing operators needed to explain the neutrino textures, while $N'$ serves as the dark matter candidate. Alternatively $N_e$ could be thought of as comparable to $N_\mu$ and $N_\tau$ with $N'_e$ heavy. In this scenario $N'_e$ would generate the appropriate mass-inducing operators. 

$L_\mu-L_\tau$ initially appeared to lend itself as a minimalistic implementation of the Shuve-Yavin progenitor scenario. It had a naturally sterile state, and---to a crude approximation---seemed to naturally produce acceptable neutrino textures. However to adequately explain neutrino phenomenology, and the observed dark matter abundance one is forced to introduce two right-handed fields beyond the minimal three if one wishes to have a theory that does not depend on irrelevant operators (dimensionality greater than four).  

\subsection{The cases of gauged $L_e-L_\mu$ and $L_e-L_\tau$}
In the limit in which the sterile fermion ($N_\tau$ and $N_\mu$ respectively) is totally decoupled it is obvious that the correct neutrino phenomenology cannot be recovered. Ergo, one must also introduce an additional sterile state into these theories to serve as the dark matter candidate. 

Analyzing these symmetries can be done using all of the machinery that was developed for $L_\mu-L_\tau$ by just interchanging the appropriate lepton indices. For example in the case of $L_\tau-L_e$ the condition $\Omega_{12}^2=\Omega_{22}$ becomes $\Omega_{12}^2=\Omega_{11}$. The analysis we performed shows that $L_e-L_\mu$ and $L_e-L_\tau$ are similar to $L_\mu-L_\tau$ in that they cannot reproduce neutrino textures in the $3N$-extension and require one additional degree of freedom. This suggests that there is no aesthetic reason to prefer the symmetries $L_e-L_\mu$ or $L_e-L_\tau$ over $L_\mu-L_\tau$, or vice-versa. 

\subsection{The case of a gauged $B-L$}
The case of a $B-L$ Shuve-Yavin progenitor scenario initially seemed to have two features that distinguished it from the gauged lepton flavour symmetries $L_i-L_j$  
\begin{enumerate}[label=(\roman*)]
\item We we forced to introduce an additional sterile fermion that did not belong to any existing lepton generations. This seemed ad-hoc.
\item To couple this new fermion to the active leptons via mass mixing we needed to charge the order parameter in such a way that the three right-handed neutrinos charged under $B-L$ did not couple to one another via mass-mixing 
\end{enumerate}
The first of these features is now clearly one that is present for any model of sterile neutrino dark matter. The second should be further elucidated. 

Suppose, in addition to the $3N$-extension, we add one additional sterile state, $N'$,  which is uncharged under $B-L$ and has a Majorana mass of $M'$. If we wish to have terms like $M_D N'N_i$ we must charge our order parameter so that $Q(S)=-Q(N_i)$ to allow terms in the Lagrangian such as $g_iSN'N_i$. Then in analogy with $\nu_{\text{\tiny{$\Sigma$}}}$ we may define $N_{\text{\tiny{$\Sigma$}}}=\frac{1}{\mathcal{N}}\left(g_e N_e+g_\mu N_\mu +g_\tau N_\tau\right)$ where the normalization is defined as $\mathcal{N}=\sqrt{g_e^2+g_\mu^2+g_\tau^2}$. Then our mixing is controlled by $\theta=\mathcal{N}v/M'$. This can be arbitrarily small as far as neutrino phenomenology is concerned because our dark matter candidate does not contribute appreciably to the generation of neutrino masses (see Section \ref{DMandneutrinos}).

Our order parameter was chosen to have charge equal and opposite to the right-handed fields. $B-L$ is generation blind and so any combination of two right-handed fields has charge two (i.e. $Q(N_iN_j)=-2Q(S)$).  So if we truncate our Lagrangian at operators of dimension four then the right-handed mass matrix vanishes. It should be noted that by integrating out $N'$ a right-handed mass matrix is produced. However as was previously argued the characteristic mass scale will be about ten millionths the observed neutrino mass scale and so this is totally negligible. 

As was noted in Section \ref{naivety} the Yukawa matrix for $B-L$ is a general $3\times3$ matrix and contains enough degrees of freedom to fit to any neutrino texture. Due to the vanishing Majorana mass matrix, this charge assignment predicts Dirac neutrinos, which means that our model provides no ``natural'' explanation for the vast disparity in masses of the neutrinos and other leptons. The model does yield a dark matter candidate, and viable neutrino phenomenology with the inclusion of only one additional sterile state beyond the $3N$-extension. This makes the model more minimalistic, in some sense, than the case of a gauged $L_\mu-L_\tau$, which required two additional sterile states.  
\section{Conclusions}\label{sec:conc}
We have investigated the implications of an Abelian $Z'$ for neutrino phenomenology and dark matter. We studied a minimal extension of the Standard Model that included a $Z'$, an order parameter necessitated by cosmological considerations, and three right-handed neutrinos labelled by the conventional lepton numbers.

We found that lepton flavour symmetries, coupled with a restriction to three right-handed states can fit to four of the five neutrino oscillation parameters such that they agree with experiment.  Tension between our model and experiment for the final parameter can be reduced to around $25\%$ and this is perhaps suggestive that some small perturbation on this model could reproduce the correct neutrino textures. 

By including higher-dimensional operators we were able to reproduce all of the mixing parameters' best-fit values. This is suggestive of a need for a fourth sterile state if one wishes to gauge lepton flavour symmetries and explain neutrino oscillations via sterile neutrinos. 

We found that for $B-L$, neutrino textures could be fit trivially with either Dirac or Majorana neutrinos due to the totally unconstrained Yukawa matrix. The latter requires an order parameter, however if one is interested in models of Abelian $Z'$s this is independently motivated due to the necessity of a massive $Z'$ to satisfy bounds coming from $N_{eff}$ \cite{Ade:2015xua}. 

We also found a trivial parametric relation that to the author's knowledge had not been emphasized in the literature before. If one proposes a model of sterile neutrino dark matter one has, in the see-saw limit, $m_\nu^{(s)}\thicksim M_s \sin^2{\theta}$ where $m_\nu^{(s)}$ is the dark matter candidate's contribution to neutrino masses. The bounds on $\sin^2{\theta}$, and the mass range for sterile neutrinos to represent a viable dark matter candidate demand that the dark matter candidate not be involved in the generation of neutrino masses beyond about ten part per million. This implies that the existence of neutrino masses does not motivate a dark matter candidate because if such a sterile neutrino candidate existed it would not contribute to the observed phenomenon appreciably. This relation suggests that any model of sterile neutrino dark matter must include an additional decoupled state beyond those involved in neutrino mass generation. 

In summary we found that for the case of gauged lepton flavour symmetries, $L_i-L_j$ four sterile neutrinos are needed to reproduce oscillation data, and a fifth is required to provide a viable dark matter candidate. The $3N$-extension with a gauged $B-L$ can produce neutrino phenomenology that agrees with experimental observations, but for the model to include a viable dark matter candidate one additional sterile state must be included. This results in Dirac neutrinos and a Majorana neutrino dark matter candidate, if it seeks to simultaneously explain neutrino oscillation data. 

\section*{Acknowledgements}
I would like to thank both Itay Yavin and Brian Shuve for their helpful discussions as well as Mathew Williams, Gabriel Magill, and Sedigh Ghamari. This research was support by funds from the National Science and Engineering Research Council of Canada, the Ontario Graduate Scholarship (OGS) program, and the Early Research Awards program of Ontario.

\bibliography{bibliothesis}

\begin{thebibliography}{37}%
\makeatletter
\providecommand \@ifxundefined [1]{%
 \@ifx{#1\undefined}
}%
\providecommand \@ifnum [1]{%
 \ifnum #1\expandafter \@firstoftwo
 \else \expandafter \@secondoftwo
 \fi
}%
\providecommand \@ifx [1]{%
 \ifx #1\expandafter \@firstoftwo
 \else \expandafter \@secondoftwo
 \fi
}%
\providecommand \natexlab [1]{#1}%
\providecommand \enquote  [1]{``#1''}%
\providecommand \bibnamefont  [1]{#1}%
\providecommand \bibfnamefont [1]{#1}%
\providecommand \citenamefont [1]{#1}%
\providecommand \href@noop [0]{\@secondoftwo}%
\providecommand \href [0]{\begingroup \@sanitize@url \@href}%
\providecommand \@href[1]{\@@startlink{#1}\@@href}%
\providecommand \@@href[1]{\endgroup#1\@@endlink}%
\providecommand \@sanitize@url [0]{\catcode `\\12\catcode `\$12\catcode
  `\&12\catcode `\#12\catcode `\^12\catcode `\_12\catcode `\%12\relax}%
\providecommand \@@startlink[1]{}%
\providecommand \@@endlink[0]{}%
\providecommand \url  [0]{\begingroup\@sanitize@url \@url }%
\providecommand \@url [1]{\endgroup\@href {#1}{\urlprefix }}%
\providecommand \urlprefix  [0]{URL }%
\providecommand \Eprint [0]{\href }%
\providecommand \doibase [0]{http://dx.doi.org/}%
\providecommand \selectlanguage [0]{\@gobble}%
\providecommand \bibinfo  [0]{\@secondoftwo}%
\providecommand \bibfield  [0]{\@secondoftwo}%
\providecommand \translation [1]{[#1]}%
\providecommand \BibitemOpen [0]{}%
\providecommand \bibitemStop [0]{}%
\providecommand \bibitemNoStop [0]{.\EOS\space}%
\providecommand \EOS [0]{\spacefactor3000\relax}%
\providecommand \BibitemShut  [1]{\csname bibitem#1\endcsname}%
\let\auto@bib@innerbib\@empty
\bibitem [{\citenamefont {Baek}\ and\ \citenamefont {Ko}(2009)}]{Baek:2008nz}%
  \BibitemOpen
  \bibfield  {author} {\bibinfo {author} {\bibfnamefont {S.}~\bibnamefont
  {Baek}}\ and\ \bibinfo {author} {\bibfnamefont {P.}~\bibnamefont {Ko}},\
  }\href {\doibase 10.1088/1475-7516/2009/10/011} {\bibfield  {journal}
  {\bibinfo  {journal} {JCAP}\ }\textbf {\bibinfo {volume} {0910}},\ \bibinfo
  {pages} {011} (\bibinfo {year} {2009})},\ \Eprint
  {http://arxiv.org/abs/0811.1646} {arXiv:0811.1646 [hep-ph]} \BibitemShut
  {NoStop}%
\bibitem [{\citenamefont {Langacker}(2009)}]{Langacker:2008yv}%
  \BibitemOpen
  \bibfield  {author} {\bibinfo {author} {\bibfnamefont {P.}~\bibnamefont
  {Langacker}},\ }\href {\doibase 10.1103/RevModPhys.81.1199} {\bibfield
  {journal} {\bibinfo  {journal} {Rev.Mod.Phys.}\ }\textbf {\bibinfo {volume}
  {81}},\ \bibinfo {pages} {1199} (\bibinfo {year} {2009})},\ \Eprint
  {http://arxiv.org/abs/0801.1345} {arXiv:0801.1345 [hep-ph]} \BibitemShut
  {NoStop}%
\bibitem [{\citenamefont {Heeck}\ and\ \citenamefont
  {Rodejohann}(2011)}]{Heeck:2011wj}%
  \BibitemOpen
  \bibfield  {author} {\bibinfo {author} {\bibfnamefont {J.}~\bibnamefont
  {Heeck}}\ and\ \bibinfo {author} {\bibfnamefont {W.}~\bibnamefont
  {Rodejohann}},\ }\href {\doibase 10.1103/PhysRevD.84.075007} {\bibfield
  {journal} {\bibinfo  {journal} {Phys.Rev.}\ }\textbf {\bibinfo {volume}
  {D84}},\ \bibinfo {pages} {075007} (\bibinfo {year} {2011})},\ \Eprint
  {http://arxiv.org/abs/1107.5238} {arXiv:1107.5238 [hep-ph]} \BibitemShut
  {NoStop}%
\bibitem [{\citenamefont {Williams}\ \emph {et~al.}(2011)\citenamefont
  {Williams}, \citenamefont {Burgess}, \citenamefont {Maharana},\ and\
  \citenamefont {Quevedo}}]{Williams:2011qb}%
  \BibitemOpen
  \bibfield  {author} {\bibinfo {author} {\bibfnamefont {M.}~\bibnamefont
  {Williams}}, \bibinfo {author} {\bibfnamefont {C.}~\bibnamefont {Burgess}},
  \bibinfo {author} {\bibfnamefont {A.}~\bibnamefont {Maharana}}, \ and\
  \bibinfo {author} {\bibfnamefont {F.}~\bibnamefont {Quevedo}},\ }\href
  {\doibase 10.1007/JHEP08(2011)106} {\bibfield  {journal} {\bibinfo  {journal}
  {JHEP}\ }\textbf {\bibinfo {volume} {1108}},\ \bibinfo {pages} {106}
  (\bibinfo {year} {2011})},\ \Eprint {http://arxiv.org/abs/1103.4556}
  {arXiv:1103.4556 [hep-ph]} \BibitemShut {NoStop}%
\bibitem [{\citenamefont {Harigaya}\ \emph {et~al.}(2014)\citenamefont
  {Harigaya}, \citenamefont {Igari}, \citenamefont {Nojiri}, \citenamefont
  {Takeuchi},\ and\ \citenamefont {Tobe}}]{Harigaya:2013twa}%
  \BibitemOpen
  \bibfield  {author} {\bibinfo {author} {\bibfnamefont {K.}~\bibnamefont
  {Harigaya}}, \bibinfo {author} {\bibfnamefont {T.}~\bibnamefont {Igari}},
  \bibinfo {author} {\bibfnamefont {M.~M.}\ \bibnamefont {Nojiri}}, \bibinfo
  {author} {\bibfnamefont {M.}~\bibnamefont {Takeuchi}}, \ and\ \bibinfo
  {author} {\bibfnamefont {K.}~\bibnamefont {Tobe}},\ }\href {\doibase
  10.1007/JHEP03(2014)105} {\bibfield  {journal} {\bibinfo  {journal} {JHEP}\
  }\textbf {\bibinfo {volume} {03}},\ \bibinfo {pages} {105} (\bibinfo {year}
  {2014})},\ \Eprint {http://arxiv.org/abs/1311.0870} {arXiv:1311.0870
  [hep-ph]} \BibitemShut {NoStop}%
\bibitem [{\citenamefont {Shuve}\ and\ \citenamefont
  {Yavin}(2014)}]{Shuve:2014doa}%
  \BibitemOpen
  \bibfield  {author} {\bibinfo {author} {\bibfnamefont {B.}~\bibnamefont
  {Shuve}}\ and\ \bibinfo {author} {\bibfnamefont {I.}~\bibnamefont {Yavin}},\
  }\href {\doibase 10.1103/PhysRevD.89.113004} {\bibfield  {journal} {\bibinfo
  {journal} {Phys.Rev.}\ }\textbf {\bibinfo {volume} {D89}},\ \bibinfo {pages}
  {113004} (\bibinfo {year} {2014})},\ \Eprint {http://arxiv.org/abs/1403.2727}
  {arXiv:1403.2727 [hep-ph]} \BibitemShut {NoStop}%
\bibitem [{\citenamefont {Foot}(1991)}]{Foot:1990mn}%
  \BibitemOpen
  \bibfield  {author} {\bibinfo {author} {\bibfnamefont {R.}~\bibnamefont
  {Foot}},\ }\href {\doibase 10.1142/S0217732391000543} {\bibfield  {journal}
  {\bibinfo  {journal} {Mod. Phys. Lett.}\ }\textbf {\bibinfo {volume} {A6}},\
  \bibinfo {pages} {527} (\bibinfo {year} {1991})}\BibitemShut {NoStop}%
\bibitem [{\citenamefont {He}\ \emph {et~al.}(1991)\citenamefont {He},
  \citenamefont {Joshi}, \citenamefont {Lew},\ and\ \citenamefont
  {Volkas}}]{He:1991qd}%
  \BibitemOpen
  \bibfield  {author} {\bibinfo {author} {\bibfnamefont {X.-G.}\ \bibnamefont
  {He}}, \bibinfo {author} {\bibfnamefont {G.~C.}\ \bibnamefont {Joshi}},
  \bibinfo {author} {\bibfnamefont {H.}~\bibnamefont {Lew}}, \ and\ \bibinfo
  {author} {\bibfnamefont {R.}~\bibnamefont {Volkas}},\ }\href {\doibase
  10.1103/PhysRevD.44.2118} {\bibfield  {journal} {\bibinfo  {journal}
  {Phys.Rev.}\ }\textbf {\bibinfo {volume} {D44}},\ \bibinfo {pages} {2118}
  (\bibinfo {year} {1991})}\BibitemShut {NoStop}%
\bibitem [{\citenamefont {Weinberg}(1995)}]{WeinbergVol2}%
  \BibitemOpen
  \bibfield  {author} {\bibinfo {author} {\bibfnamefont {S.}~\bibnamefont
  {Weinberg}},\ }\href@noop {} {\emph {\bibinfo {title} {Quantum Theory of
  Fields}}}\ (\bibinfo  {publisher} {Cambridge University Press},\ \bibinfo
  {address} {Cambridge, USA},\ \bibinfo {year} {1995})\BibitemShut {NoStop}%
\bibitem [{\citenamefont {Ahmad}\ \emph {et~al.}(2002)\citenamefont {Ahmad}
  \emph {et~al.}}]{Ahmad:2002jz}%
  \BibitemOpen
  \bibfield  {author} {\bibinfo {author} {\bibfnamefont {Q.}~\bibnamefont
  {Ahmad}} \emph {et~al.} (\bibinfo {collaboration} {SNO}),\ }\href {\doibase
  10.1103/PhysRevLett.89.011301} {\ \textbf {\bibinfo {volume} {89}},\ \bibinfo
  {pages} {011301} (\bibinfo {year} {2002})},\ \Eprint
  {http://arxiv.org/abs/nucl-ex/0204008} {arXiv:nucl-ex/0204008 [nucl-ex]}
  \BibitemShut {NoStop}%
\bibitem [{\citenamefont {Wendell}\ \emph {et~al.}(2010)\citenamefont {Wendell}
  \emph {et~al.}}]{Wendell:2010md}%
  \BibitemOpen
  \bibfield  {author} {\bibinfo {author} {\bibfnamefont {R.}~\bibnamefont
  {Wendell}} \emph {et~al.} (\bibinfo {collaboration} {Super-Kamiokande}),\
  }\href {\doibase 10.1103/PhysRevD.81.092004} {\bibfield  {journal} {\bibinfo
  {journal} {Phys.Rev.}\ }\textbf {\bibinfo {volume} {D81}},\ \bibinfo {pages}
  {092004} (\bibinfo {year} {2010})},\ \Eprint {http://arxiv.org/abs/1002.3471}
  {arXiv:1002.3471 [hep-ex]} \BibitemShut {NoStop}%
\bibitem [{\citenamefont {Abe}\ \emph {et~al.}(2011)\citenamefont {Abe} \emph
  {et~al.}}]{Abe:2010hy}%
  \BibitemOpen
  \bibfield  {author} {\bibinfo {author} {\bibfnamefont {K.}~\bibnamefont
  {Abe}} \emph {et~al.} (\bibinfo {collaboration} {Super-Kamiokande}),\ }\href
  {\doibase 10.1103/PhysRevD.83.052010} {\bibfield  {journal} {\bibinfo
  {journal} {Phys.Rev.}\ }\textbf {\bibinfo {volume} {D83}},\ \bibinfo {pages}
  {052010} (\bibinfo {year} {2011})},\ \Eprint {http://arxiv.org/abs/1010.0118}
  {arXiv:1010.0118 [hep-ex]} \BibitemShut {NoStop}%
\bibitem [{\citenamefont {Aharmim}\ \emph {et~al.}(2013)\citenamefont {Aharmim}
  \emph {et~al.}}]{Aharmim:2011yq}%
  \BibitemOpen
  \bibfield  {author} {\bibinfo {author} {\bibfnamefont {B.}~\bibnamefont
  {Aharmim}} \emph {et~al.} (\bibinfo {collaboration} {SNO}),\ }\href {\doibase
  10.1103/PhysRevC.87.015502} {\bibfield  {journal} {\bibinfo  {journal}
  {Phys.Rev.}\ }\textbf {\bibinfo {volume} {C87}},\ \bibinfo {pages} {015502}
  (\bibinfo {year} {2013})},\ \Eprint {http://arxiv.org/abs/1107.2901}
  {arXiv:1107.2901 [nucl-ex]} \BibitemShut {NoStop}%
\bibitem [{\citenamefont {Zhan}(2015)}]{Zhan:2015aha}%
  \BibitemOpen
  \bibfield  {author} {\bibinfo {author} {\bibfnamefont {L.}~\bibnamefont
  {Zhan}} (\bibinfo {collaboration} {Daya Bay}),\ }\href@noop {} {\  (\bibinfo
  {year} {2015})},\ \Eprint {http://arxiv.org/abs/1506.01149} {arXiv:1506.01149
  [hep-ex]} \BibitemShut {NoStop}%
\bibitem [{\citenamefont {Weinberg}(1979)}]{Weinberg:1979sa}%
  \BibitemOpen
  \bibfield  {author} {\bibinfo {author} {\bibfnamefont {S.}~\bibnamefont
  {Weinberg}},\ }\href {\doibase 10.1103/PhysRevLett.43.1566} {\bibfield
  {journal} {\bibinfo  {journal} {Phys.Rev.Lett.}\ }\textbf {\bibinfo {volume}
  {43}},\ \bibinfo {pages} {1566} (\bibinfo {year} {1979})}\BibitemShut
  {NoStop}%
\bibitem [{\citenamefont {D'Angelo}\ \emph {et~al.}(2014)\citenamefont
  {D'Angelo} \emph {et~al.}}]{D'Angelo:2014vgk}%
  \BibitemOpen
  \bibfield  {author} {\bibinfo {author} {\bibfnamefont {D.}~\bibnamefont
  {D'Angelo}} \emph {et~al.} (\bibinfo {collaboration} {Borexino}),\
  }\href@noop {} {\  (\bibinfo {year} {2014})},\ \Eprint
  {http://arxiv.org/abs/1405.7919} {arXiv:1405.7919 [hep-ex]} \BibitemShut
  {NoStop}%
\bibitem [{\citenamefont {Bjorken}\ \emph {et~al.}(2009)\citenamefont
  {Bjorken}, \citenamefont {Essig}, \citenamefont {Schuster},\ and\
  \citenamefont {Toro}}]{Bjorken:2009mm}%
  \BibitemOpen
  \bibfield  {author} {\bibinfo {author} {\bibfnamefont {J.~D.}\ \bibnamefont
  {Bjorken}}, \bibinfo {author} {\bibfnamefont {R.}~\bibnamefont {Essig}},
  \bibinfo {author} {\bibfnamefont {P.}~\bibnamefont {Schuster}}, \ and\
  \bibinfo {author} {\bibfnamefont {N.}~\bibnamefont {Toro}},\ }\href {\doibase
  10.1103/PhysRevD.80.075018} {\bibfield  {journal} {\bibinfo  {journal} {Phys.
  Rev.}\ }\textbf {\bibinfo {volume} {D80}},\ \bibinfo {pages} {075018}
  (\bibinfo {year} {2009})},\ \Eprint {http://arxiv.org/abs/0906.0580}
  {arXiv:0906.0580 [hep-ph]} \BibitemShut {NoStop}%
\bibitem [{\citenamefont {Essig}\ \emph {et~al.}(2009)\citenamefont {Essig},
  \citenamefont {Schuster},\ and\ \citenamefont {Toro}}]{Essig:2009nc}%
  \BibitemOpen
  \bibfield  {author} {\bibinfo {author} {\bibfnamefont {R.}~\bibnamefont
  {Essig}}, \bibinfo {author} {\bibfnamefont {P.}~\bibnamefont {Schuster}}, \
  and\ \bibinfo {author} {\bibfnamefont {N.}~\bibnamefont {Toro}},\ }\href
  {\doibase 10.1103/PhysRevD.80.015003} {\bibfield  {journal} {\bibinfo
  {journal} {Phys. Rev.}\ }\textbf {\bibinfo {volume} {D80}},\ \bibinfo {pages}
  {015003} (\bibinfo {year} {2009})},\ \Eprint {http://arxiv.org/abs/0903.3941}
  {arXiv:0903.3941 [hep-ph]} \BibitemShut {NoStop}%
\bibitem [{\citenamefont {Altmannshofer}\ \emph {et~al.}(2014)\citenamefont
  {Altmannshofer}, \citenamefont {Gori}, \citenamefont {Pospelov},\ and\
  \citenamefont {Yavin}}]{Altmannshofer:2014pba}%
  \BibitemOpen
  \bibfield  {author} {\bibinfo {author} {\bibfnamefont {W.}~\bibnamefont
  {Altmannshofer}}, \bibinfo {author} {\bibfnamefont {S.}~\bibnamefont {Gori}},
  \bibinfo {author} {\bibfnamefont {M.}~\bibnamefont {Pospelov}}, \ and\
  \bibinfo {author} {\bibfnamefont {I.}~\bibnamefont {Yavin}},\ }\href
  {\doibase 10.1103/PhysRevLett.113.091801} {\bibfield  {journal} {\bibinfo
  {journal} {Phys.Rev.Lett.}\ }\textbf {\bibinfo {volume} {113}},\ \bibinfo
  {pages} {091801} (\bibinfo {year} {2014})},\ \Eprint
  {http://arxiv.org/abs/1406.2332} {arXiv:1406.2332 [hep-ph]} \BibitemShut
  {NoStop}%
\bibitem [{\citenamefont {Lesgourgues}\ and\ \citenamefont
  {Pastor}(2006)}]{Lesgourgues:2006nd}%
  \BibitemOpen
  \bibfield  {author} {\bibinfo {author} {\bibfnamefont {J.}~\bibnamefont
  {Lesgourgues}}\ and\ \bibinfo {author} {\bibfnamefont {S.}~\bibnamefont
  {Pastor}},\ }\href {\doibase 10.1016/j.physrep.2006.04.001} {\bibfield
  {journal} {\bibinfo  {journal} {Phys. Rept.}\ }\textbf {\bibinfo {volume}
  {429}},\ \bibinfo {pages} {307} (\bibinfo {year} {2006})},\ \Eprint
  {http://arxiv.org/abs/astro-ph/0603494} {arXiv:astro-ph/0603494 [astro-ph]}
  \BibitemShut {NoStop}%
\bibitem [{\citenamefont {Ade}\ \emph {et~al.}(2015)\citenamefont {Ade} \emph
  {et~al.}}]{Ade:2015xua}%
  \BibitemOpen
  \bibfield  {author} {\bibinfo {author} {\bibfnamefont {P.}~\bibnamefont
  {Ade}} \emph {et~al.} (\bibinfo {collaboration} {Planck}),\ }\href {\doibase
  10.1051/0004-6361/201321591} {\bibfield  {journal} {\bibinfo  {journal}
  {Astron.Astrophys.}\ }\textbf {\bibinfo {volume} {571}},\ \bibinfo {pages}
  {A16} (\bibinfo {year} {2015})},\ \Eprint {http://arxiv.org/abs/1502.01589}
  {arXiv:1502.01589 [astro-ph.CO]} \BibitemShut {NoStop}%
\bibitem [{\citenamefont {Pontecorvo}(1968)}]{Pontecorvo:1967fh}%
  \BibitemOpen
  \bibfield  {author} {\bibinfo {author} {\bibfnamefont {B.}~\bibnamefont
  {Pontecorvo}},\ }\href@noop {} {\bibfield  {journal} {\bibinfo  {journal}
  {Sov.Phys.JETP}\ }\textbf {\bibinfo {volume} {26}},\ \bibinfo {pages} {984}
  (\bibinfo {year} {1968})}\BibitemShut {NoStop}%
\bibitem [{\citenamefont {Kusenko}(2009)}]{Kusenko:2009up}%
  \BibitemOpen
  \bibfield  {author} {\bibinfo {author} {\bibfnamefont {A.}~\bibnamefont
  {Kusenko}},\ }\href {\doibase 10.1016/j.physrep.2009.07.004} {\bibfield
  {journal} {\bibinfo  {journal} {Phys.Rept.}\ }\textbf {\bibinfo {volume}
  {481}},\ \bibinfo {pages} {1} (\bibinfo {year} {2009})},\ \Eprint
  {http://arxiv.org/abs/0906.2968} {arXiv:0906.2968 [hep-ph]} \BibitemShut
  {NoStop}%
\bibitem [{\citenamefont {Asaka}\ and\ \citenamefont
  {Shaposhnikov}(2005)}]{Asaka:2005pn}%
  \BibitemOpen
  \bibfield  {author} {\bibinfo {author} {\bibfnamefont {T.}~\bibnamefont
  {Asaka}}\ and\ \bibinfo {author} {\bibfnamefont {M.}~\bibnamefont
  {Shaposhnikov}},\ }\href {\doibase 10.1016/j.physletb.2005.06.020} {\bibfield
   {journal} {\bibinfo  {journal} {Phys. Lett.}\ }\textbf {\bibinfo {volume}
  {B620}},\ \bibinfo {pages} {17} (\bibinfo {year} {2005})},\ \Eprint
  {http://arxiv.org/abs/hep-ph/0505013} {arXiv:hep-ph/0505013 [hep-ph]}
  \BibitemShut {NoStop}%
\bibitem [{\citenamefont {Asaka}\ \emph {et~al.}(2005)\citenamefont {Asaka},
  \citenamefont {Blanchet},\ and\ \citenamefont {Shaposhnikov}}]{Asaka:2005an}%
  \BibitemOpen
  \bibfield  {author} {\bibinfo {author} {\bibfnamefont {T.}~\bibnamefont
  {Asaka}}, \bibinfo {author} {\bibfnamefont {S.}~\bibnamefont {Blanchet}}, \
  and\ \bibinfo {author} {\bibfnamefont {M.}~\bibnamefont {Shaposhnikov}},\
  }\href {\doibase 10.1016/j.physletb.2005.09.070} {\bibfield  {journal}
  {\bibinfo  {journal} {Phys. Lett.}\ }\textbf {\bibinfo {volume} {B631}},\
  \bibinfo {pages} {151} (\bibinfo {year} {2005})},\ \Eprint
  {http://arxiv.org/abs/hep-ph/0503065} {arXiv:hep-ph/0503065 [hep-ph]}
  \BibitemShut {NoStop}%
\bibitem [{\citenamefont {Dodelson}\ and\ \citenamefont
  {Widrow}(1994)}]{Dodelson:1993je}%
  \BibitemOpen
  \bibfield  {author} {\bibinfo {author} {\bibfnamefont {S.}~\bibnamefont
  {Dodelson}}\ and\ \bibinfo {author} {\bibfnamefont {L.~M.}\ \bibnamefont
  {Widrow}},\ }\href {\doibase 10.1103/PhysRevLett.72.17} {\bibfield  {journal}
  {\bibinfo  {journal} {Phys.Rev.Lett.}\ }\textbf {\bibinfo {volume} {72}},\
  \bibinfo {pages} {17} (\bibinfo {year} {1994})},\ \Eprint
  {http://arxiv.org/abs/hep-ph/9303287} {arXiv:hep-ph/9303287 [hep-ph]}
  \BibitemShut {NoStop}%
\bibitem [{\citenamefont {Dolgov}\ and\ \citenamefont
  {Hansen}(2002)}]{Dolgov:2000ew}%
  \BibitemOpen
  \bibfield  {author} {\bibinfo {author} {\bibfnamefont {A.}~\bibnamefont
  {Dolgov}}\ and\ \bibinfo {author} {\bibfnamefont {S.}~\bibnamefont
  {Hansen}},\ }\href {\doibase 10.1016/S0927-6505(01)00115-3} {\bibfield
  {journal} {\bibinfo  {journal} {Astropart.Phys.}\ }\textbf {\bibinfo {volume}
  {16}},\ \bibinfo {pages} {339} (\bibinfo {year} {2002})},\ \Eprint
  {http://arxiv.org/abs/hep-ph/0009083} {arXiv:hep-ph/0009083 [hep-ph]}
  \BibitemShut {NoStop}%
\bibitem [{\citenamefont {Wu}\ \emph {et~al.}(2009)\citenamefont {Wu},
  \citenamefont {Ho},\ and\ \citenamefont {Boyanovsky}}]{Wu:2009yr}%
  \BibitemOpen
  \bibfield  {author} {\bibinfo {author} {\bibfnamefont {J.}~\bibnamefont
  {Wu}}, \bibinfo {author} {\bibfnamefont {C.-M.}\ \bibnamefont {Ho}}, \ and\
  \bibinfo {author} {\bibfnamefont {D.}~\bibnamefont {Boyanovsky}},\ }\href
  {\doibase 10.1103/PhysRevD.80.103511} {\bibfield  {journal} {\bibinfo
  {journal} {Phys.Rev.}\ }\textbf {\bibinfo {volume} {D80}},\ \bibinfo {pages}
  {103511} (\bibinfo {year} {2009})},\ \Eprint {http://arxiv.org/abs/0902.4278}
  {arXiv:0902.4278 [hep-ph]} \BibitemShut {NoStop}%
\bibitem [{\citenamefont {Boyarsky}\ \emph {et~al.}(2007)\citenamefont
  {Boyarsky}, \citenamefont {Nevalainen},\ and\ \citenamefont
  {Ruchayskiy}}]{Boyarsky:2006ag}%
  \BibitemOpen
  \bibfield  {author} {\bibinfo {author} {\bibfnamefont {A.}~\bibnamefont
  {Boyarsky}}, \bibinfo {author} {\bibfnamefont {J.}~\bibnamefont
  {Nevalainen}}, \ and\ \bibinfo {author} {\bibfnamefont {O.}~\bibnamefont
  {Ruchayskiy}},\ }\href {\doibase 10.1051/0004-6361:20066774} {\bibfield
  {journal} {\bibinfo  {journal} {Astron.Astrophys.}\ }\textbf {\bibinfo
  {volume} {471}},\ \bibinfo {pages} {51} (\bibinfo {year} {2007})},\ \Eprint
  {http://arxiv.org/abs/astro-ph/0610961} {arXiv:astro-ph/0610961 [astro-ph]}
  \BibitemShut {NoStop}%
\bibitem [{\citenamefont {{Watson}}\ \emph {et~al.}(2012)\citenamefont
  {{Watson}}, \citenamefont {{Li}},\ and\ \citenamefont
  {{Polley}}}]{Watson:2012yr}%
  \BibitemOpen
  \bibfield  {author} {\bibinfo {author} {\bibfnamefont {C.~R.}\ \bibnamefont
  {{Watson}}}, \bibinfo {author} {\bibfnamefont {Z.}~\bibnamefont {{Li}}}, \
  and\ \bibinfo {author} {\bibfnamefont {N.~K.}\ \bibnamefont {{Polley}}},\
  }\href {\doibase 10.1088/1475-7516/2012/03/018} {\bibfield  {journal}
  {\bibinfo  {journal} {Journal of Cosmology and Astroparticle Physics}\
  }\textbf {\bibinfo {volume} {3}},\ \bibinfo {eid} {018} (\bibinfo {year}
  {2012})},\ \Eprint {http://arxiv.org/abs/1111.4217} {arXiv:1111.4217}
  \BibitemShut {NoStop}%
\bibitem [{\citenamefont {Notzold}\ and\ \citenamefont
  {Raffelt}(1988)}]{Notzold:1987ik}%
  \BibitemOpen
  \bibfield  {author} {\bibinfo {author} {\bibfnamefont {D.}~\bibnamefont
  {Notzold}}\ and\ \bibinfo {author} {\bibfnamefont {G.}~\bibnamefont
  {Raffelt}},\ }\href {\doibase 10.1016/0550-3213(88)90113-7} {\bibfield
  {journal} {\bibinfo  {journal} {Nucl.Phys.}\ }\textbf {\bibinfo {volume}
  {B307}},\ \bibinfo {pages} {924} (\bibinfo {year} {1988})}\BibitemShut
  {NoStop}%
\bibitem [{\citenamefont {Watson}\ \emph {et~al.}(2006)\citenamefont {Watson},
  \citenamefont {Beacom}, \citenamefont {Yuksel},\ and\ \citenamefont
  {Walker}}]{Watson:2006qb}%
  \BibitemOpen
  \bibfield  {author} {\bibinfo {author} {\bibfnamefont {C.~R.}\ \bibnamefont
  {Watson}}, \bibinfo {author} {\bibfnamefont {J.~F.}\ \bibnamefont {Beacom}},
  \bibinfo {author} {\bibfnamefont {H.}~\bibnamefont {Yuksel}}, \ and\ \bibinfo
  {author} {\bibfnamefont {T.~P.}\ \bibnamefont {Walker}},\ }\href {\doibase
  10.1103/PhysRevD.74.033009} {\bibfield  {journal} {\bibinfo  {journal}
  {Phys.Rev.}\ }\textbf {\bibinfo {volume} {D74}},\ \bibinfo {pages} {033009}
  (\bibinfo {year} {2006})},\ \Eprint {http://arxiv.org/abs/astro-ph/0605424}
  {arXiv:astro-ph/0605424 [astro-ph]} \BibitemShut {NoStop}%
\bibitem [{\citenamefont {Horiuchi}\ \emph {et~al.}(2014)\citenamefont
  {Horiuchi}, \citenamefont {Humphrey}, \citenamefont {Onorbe}, \citenamefont
  {Abazajian}, \citenamefont {Kaplinghat} \emph {et~al.}}]{Horiuchi:2013noa}%
  \BibitemOpen
  \bibfield  {author} {\bibinfo {author} {\bibfnamefont {S.}~\bibnamefont
  {Horiuchi}}, \bibinfo {author} {\bibfnamefont {P.~J.}\ \bibnamefont
  {Humphrey}}, \bibinfo {author} {\bibfnamefont {J.}~\bibnamefont {Onorbe}},
  \bibinfo {author} {\bibfnamefont {K.~N.}\ \bibnamefont {Abazajian}}, \bibinfo
  {author} {\bibfnamefont {M.}~\bibnamefont {Kaplinghat}},  \emph {et~al.},\
  }\href {\doibase 10.1103/PhysRevD.89.025017} {\bibfield  {journal} {\bibinfo
  {journal} {Phys.Rev.}\ }\textbf {\bibinfo {volume} {D89}},\ \bibinfo {pages}
  {025017} (\bibinfo {year} {2014})},\ \Eprint {http://arxiv.org/abs/1311.0282}
  {arXiv:1311.0282 [astro-ph.CO]} \BibitemShut {NoStop}%
\bibitem [{\citenamefont {Choubey}\ and\ \citenamefont
  {Rodejohann}(2005)}]{Choubey:2004hn}%
  \BibitemOpen
  \bibfield  {author} {\bibinfo {author} {\bibfnamefont {S.}~\bibnamefont
  {Choubey}}\ and\ \bibinfo {author} {\bibfnamefont {W.}~\bibnamefont
  {Rodejohann}},\ }\href {\doibase 10.1140/epjc/s2005-02133-1} {\bibfield
  {journal} {\bibinfo  {journal} {Eur. Phys. J.}\ }\textbf {\bibinfo {volume}
  {C40}},\ \bibinfo {pages} {259} (\bibinfo {year} {2005})},\ \Eprint
  {http://arxiv.org/abs/hep-ph/0411190} {arXiv:hep-ph/0411190 [hep-ph]}
  \BibitemShut {NoStop}%
\bibitem [{\citenamefont {Araki}\ \emph {et~al.}(2012)\citenamefont {Araki},
  \citenamefont {Heeck},\ and\ \citenamefont {Kubo}}]{Araki:2012ip}%
  \BibitemOpen
  \bibfield  {author} {\bibinfo {author} {\bibfnamefont {T.}~\bibnamefont
  {Araki}}, \bibinfo {author} {\bibfnamefont {J.}~\bibnamefont {Heeck}}, \ and\
  \bibinfo {author} {\bibfnamefont {J.}~\bibnamefont {Kubo}},\ }\href {\doibase
  10.1007/JHEP07(2012)083} {\bibfield  {journal} {\bibinfo  {journal} {JHEP}\
  }\textbf {\bibinfo {volume} {07}},\ \bibinfo {pages} {083} (\bibinfo {year}
  {2012})},\ \Eprint {http://arxiv.org/abs/1203.4951} {arXiv:1203.4951
  [hep-ph]} \BibitemShut {NoStop}%
\bibitem [{\citenamefont {Plestid}(2015)}]{plestid:2015thesis}%
  \BibitemOpen
  \bibfield  {author} {\bibinfo {author} {\bibfnamefont {R.}~\bibnamefont
  {Plestid}},\ }\emph {\bibinfo {title} {{The Implications of Gauged Lepton
  Flavour Symmetries for Dark Matter and Neutrino Masses}}},\ \href@noop {}
  {Master's thesis},\ \bibinfo  {school} {McMaster Universtiy} (\bibinfo {year}
  {2015}),\ \Eprint {http://arxiv.org/abs/http://hdl.handle.net/11375/18343}
  {http://hdl.handle.net/11375/18343} \BibitemShut {NoStop}%
\bibitem [{\citenamefont {Gonzalez-Garcia}\ \emph {et~al.}(2014)\citenamefont
  {Gonzalez-Garcia}, \citenamefont {Maltoni},\ and\ \citenamefont
  {Schwetz}}]{Gonzalez-Garcia:2014bfa}%
  \BibitemOpen
  \bibfield  {author} {\bibinfo {author} {\bibfnamefont {M.}~\bibnamefont
  {Gonzalez-Garcia}}, \bibinfo {author} {\bibfnamefont {M.}~\bibnamefont
  {Maltoni}}, \ and\ \bibinfo {author} {\bibfnamefont {T.}~\bibnamefont
  {Schwetz}},\ }\href {\doibase 10.1007/JHEP11(2014)052} {\bibfield  {journal}
  {\bibinfo  {journal} {JHEP}\ }\textbf {\bibinfo {volume} {1411}},\ \bibinfo
  {pages} {052} (\bibinfo {year} {2014})},\ \Eprint
  {http://arxiv.org/abs/1409.5439} {arXiv:1409.5439 [hep-ph]} \BibitemShut
  {NoStop}%
\end{thebibliography}%
\end{document}